\documentclass[fleqn,twoside]{article}
\usepackage{amssymb,amsfonts,espcrc2}
\usepackage{graphicx}

\newcommand{\ind}[2]{^{#1}_{\mbox{\scriptsize #2}}}
\newcommand{\al}[2]{\alpha\ind{#1}{#2}}
\newcommand{\sla}{\slash \hspace{-0.22cm}}
\newcommand{\slaq}{\slash \hspace{-0.18cm}}      

\def\nf{n_{\mbox{\scriptsize f}}}

\title{Quark gap equation within the analytic approach to QCD}

\author{A.C.~Aguilar\address[UV]{Departamento de F\'\i sica Te\'orica
and IFIC, Centro Mixto, Universidad de Valencia--CSIC, \\
~$\,$E-46100, Burjassot, Valencia, Spain},
A.V.~Nesterenko\addressmark[UV]\address[JINR]{Bogoliubov Laboratory
of Theoretical Physics, JINR, Dubna, 141980, Russian Federation}
and
J.~Papavassiliou\addressmark[UV]}

\begin{document}

\begin{abstract}
The compatibility between the QCD analytic invariant charge and
chiral symmetry breaking is examined in detail. The coupling in
question incorporates asymptotic freedom and infrared enhancement
into a single expression, and contains only one adjustable parameter
with dimension of mass. When inserted into the standard form of the
quark gap-equation it gives rise to solutions displaying singular
confining behavior at the origin. By relating these solutions to the
pion decay constant, a rough estimate of about $880\,$MeV is obtained
for the aforementioned mass-scale.
\vskip-1mm
\end{abstract}

\maketitle
  
     As has been advocated in recent years by a number of 
authors~\cite{ShSol}, a possible way for gaining further insight on
the infrared sector of QCD is to resort to arguments based on the
analyticity of the underlying dynamics,  as captured  by the
corresponding dispersion relations. Jointly with the renormalization
group formalism, these relations constitute the essential
ingredients  of the so-called ``analytic approach''~\cite{ShSol} to
Quantum Field Theory (some applications of this method can be 
found in Refs.~\cite{ShSol,Applics,PRD,CSBAIC}).

\begin{figure}[t]
\centerline{\includegraphics[width=75mm]{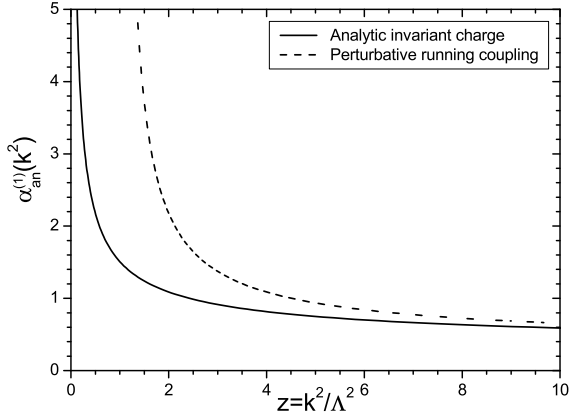}}
\vskip-7.5mm
\caption{The analytic invariant charge~(\protect\ref{AIC1L}) and the
perturbative running coupling at the one-loop level (solid and dashed
curves, respectively).}
\label{Plot:AIC}
\end{figure}

     One of the most celebrated applications of this approach to QCD
is related to the running coupling. In particular, one is able to
extrapolate the behavior of the strong coupling from the ultraviolet
region towards the infrared regime, by imposing requirements of
analyticity on it or on its derivative~\cite{ShSol,PRD}.
Specifically, if we apply such ideas on the perturbative expansion
for the one-loop $\beta$~function, the solution for the analytic
invariant charge that emerges assumes  the form~\cite{PRD}
\begin{equation}
\label{AIC1L}
\al{(1)}{an}(k^2) = \frac{4 \pi}{\beta_{0}}\, \frac{z - 1}{z \, \ln z},
\qquad z = \frac{k^2}{\Lambda^2}. 
\end{equation}
The invariant charge~(\ref{AIC1L}) has the correct analytic
properties in the $k^2$~variable, i.e., unlike its perturbative
counterpart, it displays no Landau pole. In addition, it has no
adjustable parameters other than  a mass-scale, to be denoted by
$\Lambda$, emerging, as usual, through the  standard mechanism of   
dimensional transmutation. Moreover, the coupling of~(\ref{AIC1L})
incorporates  asymptotic freedom and infrared enhancement into a 
single expression (see Figure~\ref{Plot:AIC}). Regarding this last
point, the invariant charge~(\ref{AIC1L}) has been shown to generate
the confining static quark-antiquark potential with a quasilinear
raising behavior at large distances; specifically $V(r)\simeq r/\ln
r$ when $r\to\infty$ ($r$ is the dimensionless distance between quark
and antiquark). Further appealing features of the analytic running
coupling~(\ref{AIC1L}) and its applications can be found in
Ref.~\cite{PRD}. Evidently, it would be interesting to further
scrutinize the advantages and possible limitations of the  analytic
charge in the region where it aspires to make a difference, namely
the infrared sector of QCD.

     Chiral symmetry breaking~(CSB) and dynamical mass generation are
inherently non-perturbative effects, whose study in the continuum
leads almost invariably to a treatment based on the Schwinger-Dyson
(SD) equations of the theory. As is well known to the SD experts, the
form of the solutions obtained depends crucially  on the way one
chooses to model the QCD running coupling at low energies. In this
talk we summarize results presented in~\cite{CSBAIC} on the impact of
the analytic invariant charge~\cite{PRD} on CSB. This is accomplished
by studying the solutions that emerge from  the standard gap equation
governing the quark propagator, when the aforementioned model for the
infrared behavior of the QCD coupling is adopted.

     We next proceed to the study of the gap equation. According to
the standard lore, the starting point is to express the fully dressed
quark propagator in the following general form~\cite{roberts}
\begin{equation}
S^{-1}(p) = i\sla{p} +m_0 + \Sigma(p) =
i\sla{p}A(p^2) + B(p^2).
\label{qprop}
\end{equation}
We consider the  case without explicit CSB, i.e., bare mass $m_0=0$,
and the quark mass is generated exclusively through dynamical
effects. Then CSB takes place when the self-energy $\Sigma(p)$
develops a non-zero value. Traditionally one defines the quark mass
function $M(p^2)$ in terms of the functions $A(p^2)$ and $B(p^2)$, as
$M(p^2)=B(p^2)/A(p^2)$; then CSB occurs  when $B(p^2)\neq 0$.

\begin{figure}[t]
\centerline{\includegraphics[width=75mm]{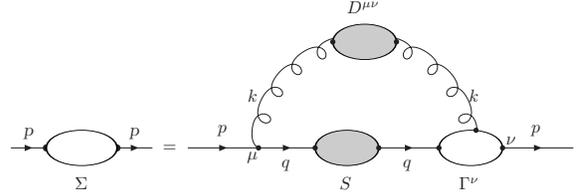}}
\vskip-7.5mm
\caption{The SD equation (\protect\ref{senergy}) for the quark
self-energy. The black blobs represent the fully dressed quark and
gluon propagators and the white one is the proper quark-gluon
vertex.} 
\label{self}
\end{figure}

     The  quark self-energy, which is represented schematically in
Figure~\ref{self}, can be written as
\begin{equation}
\label{senergy}
\Sigma(p)=\frac{4}{3}g^2\int\frac{d^4q}{(2\pi)^4}
\gamma_{\mu}S(q)\Gamma_{\nu}(q,p)D^{\mu\nu}(k),
\end{equation}
where we have used that  $\sum_a \lambda^a \lambda^a=4/3$, 
$\lambda^a$ being the Gell-Mann matrices, and $k=p-q$.  According to
this equation, the self-energy $\Sigma(p)$ is dynamically determined 
in terms of itself, the full   gluon propagator, denoted by 
$D_{\mu\nu}^{ab}(k)=\delta^{ab}D_{\mu\nu}(k)$, and the full
quark-gluon vertex~$\Gamma_{\nu}(q,p)$. Of course, both
$D_{\mu\nu}^{ab}(k)$ and $\Gamma_{\nu}(q,p)$ obey their own
complicated  SD equation, a fact which eventually makes  unavoidable
the use of  simplifications and further modeling of the unknown
functions  involved. 

     The first step toward a construction of a manageable system of
equations is  to ignore the ghost contributions in the quark SD
equation. The latter are contained in the full gluon propagator and
the full quark-gluon vertex. More specifically, the full quark-gluon
vertex  should satisfy the Slavnov-Taylor identity
\begin{eqnarray}
i\, k^{\mu}\Gamma_{\mu}(p,q)\left[1+b(k^2)\right] = &&
\nonumber \\ &&\hspace{-45mm}
=\left[1-B(k,q)\right]S^{-1}(p) - S^{-1}(q)\left[1-B(k,q)\right]\!,
\label{slavnov}
\end{eqnarray} 
where $b(k^2)$ is the ghost self-energy and $B(k,q)$ is the
ghost-quark scattering kernel. Without ghosts,
equation~(\ref{slavnov}) becomes identical to the text-book QED Ward
identity \cite{atk}, and the quark-gluon vertex now obeys 
\begin{equation}
i\, k^{\mu} \Gamma_{\mu}(p,q)=S^{-1}(p)-S^{-1}(q).
\label{ward}
\end{equation}
This last Ward identity enforces the  equality $Z_1=Z_2$ between the
vertex- and quark-wave-function renormalization  constants, exactly
as happens in QED. Evidently, in this treatment we are effectively
assuming a theory with  Abelian-like characteristics, where it is
hoped that  the form of the effective gluon propagator employed
retains some memory of the omitted ghost effects.

     The next step is to employ a ``gauge technique'' inspired
Ansatz~\cite{delbourgo} for the unknown vertex $\Gamma_{\mu}(p,q)$.
According to this technique, $\Gamma_{\mu}(p,q)$ is written in terms 
of $S(p)$ in such way as to satisfy {\it by construction} the  Ward
identity~(\ref{ward}). Clearly, this procedure introduces the
additional ambiguity of how to  fix the transverse part of the
vertex, which may lead in higher orders to mishandling of the 
overlapping divergences, but is expected to be of little consequence
in the infrared. In the following analysis we will use for 
$\Gamma_{\mu}(p,q)$ the simple Ansatz proposed in
references~\cite{atk,KTW}
\begin{eqnarray}
i\,\Gamma_{\mu}(p,q)= iA(p^2)\gamma_{\mu}+&&
\nonumber  \\ &&\hspace{-40mm}
+\frac{k_{\mu}}{k^2}\Bigl\{i\!\left[A(p^2)-A(q^2)\right]\slaq{q}
+\!\left[B(p^2)-B(q^2)\right]\!\Bigr\}. 
\label{vert}
\end{eqnarray}
Another important consequence of the   forced ``Abelianization''
imposed on the theory establishes finally the  required link between
the  gap equation and the effective QCD charge. In particular, since
now $Z_1=Z_2$,  one may define a {\it
renormalization-group-invariant  quantity},  to be denoted by
$\alpha(k^2)$, which is the exact analogue of the  QED effective
charge, namely  
\begin{equation}
g^2D_{\mu\nu}(k)=\left\{\delta_{\mu\nu} -
\frac{k_{\mu}k_{\nu}}{k^2}\right\} \frac{4\pi\alpha(k^2)}{k^2}.
\label{prop}
\end{equation}
Choosing the {\it Landau gauge} leads to the further simplification
\begin{equation}
g^2D^{\mu\nu}(k)\Gamma_{\nu}(q,p) = g^2D^{\mu\nu}(k)A(q^2)\gamma_{\nu},
\label{prod}
\end{equation}
since, in that case, the gluon propagator is completely transverse. 

     Substituting equations~(\ref{qprop}) and~(\ref{prod}) into the
quark gap  equation~(\ref{senergy}), one arrives at the commonly
used  coupled system for the  the quark self-energy in terms of 
$A(p^2)$ and $B(p^2)$~\cite{roberts,KTW}. The last necessary step in
order to set up our final quark SD equation is to employ the usual
{\it angular approximation}, which allows to handle the dependence on
the angle appearing in the arguments of the various functions
entering into the gap equation. Even though the  robustness of this 
approximation has been  occasionally put into doubt (see, for
example,~\cite{Papavassiliou:1991hx}), it offers the major advantage 
of reducing the coupled system  to one single equation, since it
automatically forces the relation $A(p^2)=1$. Therefore, we can
straightforwardly  relate $B(p^2)$ to the dynamical mass~$M(p^2)$: 
\begin{eqnarray}
\label{sde}
{\mathcal M}(x) =\frac{1}{\pi}\left[
\frac{\alpha(x)}{x}\int_0^x \frac{y {\mathcal M}(y)}
{y+{\mathcal M}^2(y)} dy + \right. \nonumber \\
\left.+ \int_x^{\infty} \frac{\alpha(y) {\mathcal M}(y)}
{y+{\mathcal M}^2(y)} dy
\right],
\end{eqnarray}
where ${\mathcal M}(x)= M(x\Lambda^2)/\Lambda$, $\,x=p^2/\Lambda^2$,
and $y=q^2/\Lambda^2$. 

     Evidently, the invariant charge $\alpha(k^2)$ lies in the heart
of the above equation. Indeed, after the succession of
simplifications listed above, it is the sole ingredient which remains
to be modeled. It is at this point that the {\it analytic} charge
$\al{(1)}{an}(k^2)$ of~(\ref{AIC1L}) will enter into the game, being
plugged into equation~(\ref{sde}) as our model for the QCD effective
charge $\alpha(k^2)$.

     Before solving the resulting integral equation numerically, we
can infer some qualitative conclusions about  the deep infrared
behavior  of the solutions  by differentiating both sides twice with
respect to $x$ (see also references~\cite{atk,KTW}), thus
converting~(\ref{sde}) into a differential equation
\begin{equation}
{\mathcal M}(v)\!
\left(\frac{d^2}{d v^2} + 2 \frac{d}{d v}\right)\!{\mathcal M}(v) =
\frac{8}{\beta_{0} v}, \;\: v \to -\infty,
\label{ndif}
\end{equation}
where $v = \ln x$. This equation may be solved iteratively; the first
iteration, corresponding to the leading infrared behavior of the
dynamical mass function ${\mathcal M}(x)$, is obtained by omitting
the second-order derivative in equation~(\ref{ndif}). Then, the
solution is 
\begin{equation}
\label{lead}
{\mathcal M}(x) \simeq \sqrt{\frac{8}{\beta_0}\ln |\ln x|}, 
\qquad x \to 0.
\end{equation}  
The above approximate solution presents a singular behavior in 
the low-energy region, which can be interpreted as a hint for
confinement (see additional discussion in
references~\cite{CSBAIC,KTW,JMC80}).

     As for the ultraviolet behavior of the solution, the
conservation of the axial-vector current eventually leads, for
sufficient large momenta, to   
\begin{equation}
{\mathcal M}(x) \simeq \frac{D}{x}(\ln x)^{\lambda -1}, \qquad
x \to \infty,
\end{equation}
where $\lambda = 12/(33-2\nf)$ is the anomalous dimension of the
mass, and $D$ is a constant independent of the renormalization point 
and directly related to the quark condensate (see also
review~\cite{roberts} and references therein).

\begin{figure}[th]
\centerline{\includegraphics[width=75mm]{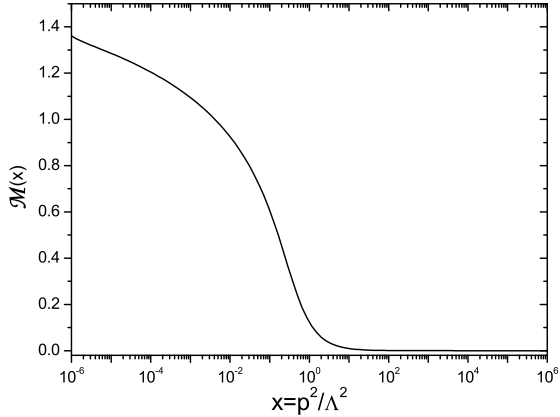}}
\vskip-7.5mm
\caption{The dimensionless quark dynamical mass ${\mathcal M}(x)$, 
obtained from Eq.~(\ref{sde}) using the one-loop analytic
charge~(\ref{AIC1L}) for $\nf=4$ active flavors.}
\label{csb-1aic}
\end{figure}

     The dynamical quark mass function ${\mathcal M}(x)$, obtained
solving numerically the integral equation~(\ref{sde}), is presented
in Figure~\ref{csb-1aic}. Indeed, one can see a soft enhancement of
${\mathcal M}(x)$ when $x \to 0$, as forecasted by Eq.~(\ref{lead}).
In order to restore a physical scale in the dimensionless mass
function ${\mathcal M}(x)$, one has to relate the obtained solution
to a QCD observable. In particular, this can be accomplished by
making use of the method developed by Pagels,
Stokar~\cite{Pagels:1979hd}, and Cornwall~\cite{JMC80}, which
provides the required link between ${\mathcal M}(x)$ and the pion
decay constant:
\begin{equation}
f_{\pi}^2=\!\frac{3\Lambda^2}{4\pi^2}\!\! \int\limits_{0}^{\infty}\! 
\left[{\mathcal M}(y)\!-\!\frac{y}{2}\frac{d{\mathcal M}(y)}{dy}\right]\!\!
\frac{y{\mathcal M}(y)\, dy}{\left[y+{\mathcal M}^2(y)\right]^2}.
\label{fpim}
\end{equation}
Thus, the scale parameter $\Lambda$ (which is the only adjustable
parameter within the approach at hand) can be evaluated by requiring
the right-hand side of Eq.~(\ref{fpim}) to acquire the experimental
value of the pion decay constant $f_{\pi}^{2}=(93\,\mbox{MeV})^{2}$.
Ultimately, for the case of $\nf=4$ active flavors, this results in
estimate $\Lambda=880\,$MeV. The higher loop corrections to the
analytic running coupling $\al{}{an}(k^2)$~\cite{PRD} do not alter
qualitatively the picture obtained. Specifically, the confining
behavior of the dynamical mass function ${\mathcal M}(x)$ persists,
but the infrared singularity displayed is weaker than that of
Eq.~(\ref{lead}). In turn, this further elevates the estimate of the
scale parameter~$\Lambda$.

     The conclusion that we draw from the above analysis is that the
analytic invariant charge developed in Ref.~\cite{PRD} and the SD
equations may coexist in a complementary and qualitatively consistent
picture, and that the obtained value for the only free parameter 
turns out to be in the right ballpark. 

\section*{Acknowledgments}

     A.N.~and J.P.~thank the organizers of QCD05 for their
hospitality. This work was supported by grants SB2003-0065 of the
Spanish Ministry of Education, CICYT FPA20002-00612, RFBR
05-01-00992, NS-2339.2003.2, and by Coordena\c{c}\~{a}o de
Aperfei\c{c}oamento de Pessoal de N\'{\i}vel Superior (Capes/Brazil)
through grant 2557/03-7 (A.C.A).

\end{document}